\documentclass[useAMS,usenatbib]{mn2e}
\usepackage{graphicx}
\usepackage[update,prepend]{epstopdf}

\title[Langmuir modes]{Pulsar emission: Langmuir modes in a relativistic multi-component plasma}
\author[P. B. Jones]{P. B Jones\thanks{E-mail:
p.jones1@physics.ox.ac.uk}  \\
University of Oxford, Department of Physics, Denys Wilkinson Building,\\
Keble Road, Oxford OX1 3RH, U.K.}

\begin{document}

\date{}

\pagerange{\pageref{firstpage}--\pageref{lastpage}} \pubyear{}

\maketitle

\label{firstpage}

\begin{abstract}

Ions, protons and possibly a small flux of electrons and positrons are accelerated outward from the polar cap of a normal or millisecond pulsar whose rotational spin is antiparallel with its magnetic moment.  The Langmuir modes of this relativistic plasma have several properties of significance for the origin of coherent radio emission.  The characteristics of the mode are determined by the sequence of singularities in the dielectric tensor at real angular frequencies, which in turn is fixed by the electron-positron momentum distribution.  We find that under a certain condition on its momentum distribution, an electron-positron flux two orders of magnitude smaller than the Goldreich-Julian flux stabilizes the plasma and extinguishes the mode.  But more generally, both the growth rate and wavenumber of the multi-component Langmuir mode can be as much as an order of magnitude larger than those of the two-component ion-proton mode.  It appears to be a further effective source for the plasma turbulence whose decay is probably responsible for the observed emission.

\end{abstract}


\section{Introduction}

Soon after the discovery of pulsars nearly half a century ago (Hewish et al 1968; see also Pacini 1967, Gold 1968), Langmuir modes were viewed as a likely source of the coherent radio emission which is their principal characteristic.  The radio-frequency power is large, and it is difficult to see that its origin can be other than the longitudinal kinetic energy of particles accelerated at the magnetic polar caps.  In principle, the plasma dielectric tensor has a very simple form: cyclotron transition rates  near the polar caps are so large that particles are in the Landau ground state and their motion is strictly one-dimensional. Modes that are precisely longitudinal do not couple directly with the radiation field, although Asseo, Pelletier \& Sol (1990) have shown that this is not true for quasi-longitudinal modes whose wave-vector ${\bf k}$ have
components $k_{\perp}\neq 0$. (In this paper, perpendicular and parallel components are defined with respect to the local magnetic flux density
${\bf B}$.)  Studies of the modes have inevitably been, at least in essence, one-dimensional.  But the physical nature of the real polar-cap plasma must be more complicated than this. Transverse dimensions are not, in all cases, very many orders of magnitude larger than mode wavelengths. There are also transverse gradients in parameters that determine the growth rate of the mode, particularly Lorentz factors, near the surface separating open from closed magnetic flux lines.  This level of complexity suggests that the existence of  modes with adequate growth rates should result in nonlinearity and the formation of a turbulent system of charge and current fluctuations which couples with radiative modes that are capable of propagating through and leaving the magnetosphere. The  hypothesis appears reasonable, but given the complexity, we are unable to offer any proof. It is distinct from other work which has considered specific mechanisms in detail, for example, maser emission due to curvature drift (Luo \& Melrose 1992) or collective curvature radiation (Kaganovitch \& Lyubarsky 2010; Istomin, Philippov \& Beskin 2012).  The ion-proton plasma Lorentz factors considered here are only moderately large and would give coherent curvature radiation only at frequencies of the order of 1 MHz.

The components of the plasma are of particular interest in the case of neutron stars whose rotational spin ${\bf \Omega}$ satisfies ${\bf \Omega}\cdot{\bf B} < 0$ at the polar caps. The special significance of this case is that the reverse flux of electrons incident on the polar cap produces protons thus giving a two-component atmosphere in local thermodynamic equilibrium (LTE) and an outward accelerated flux of ions and protons. This contrasts with the ${\bf \Omega}\cdot{\bf B} > 0$ case in which electrons are the only negatively charged particles that can be accelerated outward. For further details of this process we refer to Jones (2010). The electron flux arises from the photo-electric absorption in accelerated ions of blackbody radiation from an area of the neutron-star surface of the order of a steradian centred on the polar cap. Again, we refer to Jones (2012a,b) for calculations of the photoelectric transition rates as functions of the magnetic flux density, surface temperatures and ion atomic numbers. These processes are all quite prosaic and there are no known reasons for supposing that they do not occur. These papers assume that space-charge-limited flow boundary conditions are satisfied.

The aim of this sequence of papers has been to approach the problem of understanding the coherent radio emission of pulsars by first establishing the nature of the accelerated plasma.  To this end, a polar-cap model has been constructed (Jones 2013a) based on the calculated photo-electric transition probabilities and on the estimated diffusion time for protons to enter the LTE atmosphere.  The polar-cap model demonstrates two significant, but unsurprising, properties.  Firstly, the photo-ionization and consequent reverse flux of electrons much reduces the acceleration field, so that the outward flux of protons and ions is only moderately relativistic, certainly near the periphery of the polar cap.  With the possible exception of very young pulsars, the acceleration potential difference is so small that electron-positron formation by magnetic conversion of curvature photons cannot occur above the polar cap.  Secondly, both the  potential and the composition of the accelerated flux are time-varying, reflecting the proton diffusion time, in way that is not inconsistent with the almost universally observed pulse-to-pulse variation (see, for example, Weltevrede, Edwards \& Stappers 2006).

The composition and Lorentz factors of the outward accelerated flux lead at once to the possible role of a quasi-longitudinal Langmuir mode, following the work of Asseo et al (1990).  Investigation of this (Jones 2012b) gave growth rates adequate to produce non-linearity followed by plasma turbulence.
At the altitudes of the observed emission regions, the mode frequencies are of the same order as the low-frequency turn-overs in the radio spectra.  It is anticipated that higher wavenumbers are generated in the development of the turbulence, as is naturally the case in other turbulent systems, and consistent with the numerical studies of Weatherall (1997) on the source of very short time-scale coherent emission.  The specific example of PSR B1133+16 was considered (Jones 2013b) in relation to the properties of the ion-proton mode.  The relevance of the mode to the observed properties of the coherent emission has been described, more generally, in further papers (Jones 2014a,b). The ion-proton plasma is a direct consequence of well understood processes and it is unfortunate that these have been neglected,in nearly half a century of work, in favour of an exclusive concern with electron-positron plasmas.

However, even if curvature radiation produced electron-positron pairs are excluded, there remain possible sources of an electron-positron plasma component having number densities no more than of the order of the Goldreich-Julian density. These are considered in Section 3 and are the motivation for this paper. The smallness of the electron mass relative to that of the proton has the consequence that such a component would be represented by a numerically large term in the plasma dielectric tensor. Thus there can be at least three components contributing to the dielectric tensor. The purpose of the present paper is to give a more complete description of the ion-proton quasi-longitudinal mode than was contained in previous papers and, in particular, to investigate the effect of possible electron-positron components.

\section{The dispersion relation}

In the high magnetic-flux limit approached near the polar caps, the dispersion relation for the quasi-longitudinal mode studied by Asseo et al is determined by the $D_{zz}$ component of the dielectric tensor (see, for example, Beskin \& Philippov 2012) and is,
\begin{eqnarray}
(\omega^{2} - k_{\parallel}^{2})D_{zz} = k_{\perp}^{2},  \nonumber \\
D_{zz} = 1 - \sum_{i}\frac{\omega_{pi}^{2}}{\gamma_{i}^{3}(\omega
- k_{\parallel}v_{i})^{2}},
\end{eqnarray}
in which $v_{i}$ and $\gamma_{i}$ are the velocity (in units of c) and the Lorentz factor of the $i$-th component and the constant is $\omega_{pi}^{2} =
4\pi n_{i}q_{i}^{2}/m_{i}$, where $q_{i}$, $m_{i}$ and $n_{i}$ are, respectively its charge, mass, and number density.  Equation (1) is expressed in the corotating neutron-star frame or in the observer frame between which, close to the polar cap, we make no distinction.

Components $i=1,2$ are, respectively, ions assumed to have fixed charge $Z_{\infty}$ and protons.  Both have $\delta$-function velocity distributions.  The remaining components can represent a continuum by defining an electron-positron momentum distribution $f(p)$ normalized to unity and replacing the summation by an integral over $p$.  Integration by parts then gives the dielectric tensor component,
\begin{eqnarray}
D_{zz}= 1 -\frac{\omega_{1}^{2}}{\gamma_{1}^{3}(\omega - k_{\parallel}
v_{1})^{2}} - \frac{\omega_{2}^{2}}{\gamma_{2}^{3}(\omega - k_{\parallel}
v_{2})^{2}} +      \nonumber    \\
 \frac{m_{e}\omega_{e}^{2}}{k_{\parallel}}
\int^{\infty}_{-\infty}dp\frac{\partial f}{\partial p}\frac{1}{\omega -
k_{\parallel}v(p) + {\rm i}\epsilon},
\end{eqnarray}
in which the electron-positron term is identical with that given by Buschauer \& Benford (1977), and $\omega_{e}^{2}$ is defined in terms of the sum $n_{e}$ of the electron and positron number densities.  The infinitesimal positive quantity $\epsilon$ defines the  contribution of the singularity to the integral.  It determines the sign of the imaginary part of the dielectric tensor in terms of the gradient of the electron-positron momentum distribution, so that Im$D_{zz} > 0$ for $\partial f/\partial p < 0$.  This represents the  irreversible flow of energy from the mode to the electron-positron system (Landau damping; a positive gradient represents the reverse process).  It is not of concern here because the electron-positron distribution is itself a component of the mode and its capacity to gain energy without radical change of $\partial f/\partial p$ is very small.  In any case, the form of the electron-positron spectrum in the classical region of single-photon pair creation appears not to have been investigated (see Harding \& Lai 2006) and we shall obviate the problem by adopting a uniform distribution between limits $p_{0}$ and $p_{m}$.

Equation (2) can be simplified, as in the paper of Asseo et al, by introducing a dimensionless variable for the angular frequency of the mode,
$\omega = (1 + s)\omega_{1}^{*}$ in which $\omega_{1}^{*} = \omega_{1}\gamma_{1}^{-3/2}$, and a reference wavenumber $k_{0} = 2\gamma_{1}^{2}\omega_{1}^{*}$.
For a uniform electron-positron momentum distribution, the gradient is,
\begin{eqnarray}
\frac{\partial f}{\partial p} = \frac{1}{p_{m} - p_{0}}(\delta(p - p_{0})
- \delta(p - p_{m})),
\end{eqnarray}
and assumes that there are no reverse particles, so that $p_{0} > 0$. The dispersion relation then becomes,
\begin{eqnarray}
1 - \frac{1}{(1 + s)^{2}} - \frac{C}{(\mu + s)^{2}} - \frac{C_{e}}
{(\nu_{0} + s)(\nu_{m} + s)}  =    \nonumber    \\
\frac{k_{\perp}^{2}\gamma_{1}^{2}}
{k_{\parallel}(k_{0}(1 + s) - k_{\parallel})},
\end{eqnarray}
for the general quasi-longitudinal case. The dimensionless constants in this equation are,
\begin{eqnarray}
C & = & \frac{\omega_{2}^{*2}}{\omega_{1}^{*2}} = \frac{\gamma_{1}^{3}}
{\gamma_{2}^{3}}\frac{\alpha_{p}}{\alpha_{z}}\frac{A}{Z_{\infty}},
     \nonumber   \\
C_{e} & = & \frac{\omega_{e}^{2}(v_{m} - v_{0})}
{(\gamma_{m} -\gamma_{0})\omega_{1}^{*2}} = \frac{\alpha_{e}}{\alpha_{z}}
\frac{A}{Z_{\infty}}\frac{m_{p}}{m_{e}}\frac{\gamma_{1}^{3}
(\gamma_{m} + \gamma_{0})}{2\gamma_{0}^{2} \gamma_{m}^{2}},   \nonumber   \\
\mu & = & 1 + 2\gamma_{1}^{2}(v_{1} - v_{2})\frac{k_{\parallel}}{k_{0}}
\approx 1 - \left(1 - \frac{\gamma_{1}^{2}}{\gamma_{2}^{2}}\right)
\frac{k_{\parallel}}{k_{0}},   \nonumber  \\
\nu_{0} & = & 1 + 2\gamma_{1}^{2}(v_{1} - v_{0})\frac{k_{\parallel}}{k_{0}}
\approx 1 - \left(1 - \frac{\gamma_{1}^{2}}{\gamma_{0}^{2}}\right)
\frac{k_{\parallel}}{k_{0}},   \nonumber   \\
\nu_{m} & = & 1 + 2\gamma_{1}^{2}(v_{1} - v_{m})\frac{k_{\parallel}}{k_{0}}
\approx 1 - \left(1 - \frac{\gamma_{1}^{2}}{\gamma_{m}^{2}}\right)
\frac{k_{\parallel}}{k_{0}}.
\end{eqnarray}
The constant $C$ depends on the ion mass number $A$ and on the relative current densities of ion and proton components ($\alpha_{z}$ and $\alpha_{p}$)  expressed as fractions of the Goldreich-Julian density.  
These are $\alpha_{p}/\alpha_{z} = n_{2}/n_{1}Z_{\infty}$.  We do not distinguish between electrons and positrons here so that
$\alpha_{e}/\alpha_{z} = n_{e}/n_{1}Z_{\infty}$ and its value is not otherwise constrained.
 
We consider first the longitudinal case in which $k_{\perp} = 0$.  Equation (4) is then a sextic whose structure is transparent given its singularities on the real-$s$ axis and the simple form of the first two terms. Thus the number of real roots can usually be found by graphical inspection.  Once these have been found numerically, the complex roots are easily obtained.  We find cases of 2, 4 or 6 real roots depending on the particulars of the electron-positron distribution.  The calculated mode growth-rates given in Section 3 are for the longitudinal case. The geometrical form of the polar-cap flux-lines is such that we anticipate only quasi-longitudinal modes with $k_{\perp} \ll k_{\parallel}$.  The $k_{\perp} \neq 0$ term in equation (4) contributes a simple pole and it is easy to see that this introduces an additional real root in every case.  Graphical inspection shows it is so positioned that it should not change the broad character of our results in any significant way.

\section{The effect of an electron-positron component}

In order to examine the effect of an electron-positron component on the Langmuir mode we have first to decide on its likely number density and the limits of its momentum distribution. So it is necessary to adopt a specific polar-cap model that is capable of giving the required information about these quantities.  The polar cap chosen is that developed in previous papers (Jones 2012b, 2013a) in which photo-ionization of accelerated particles reduces electric fields to values well below those predicted using the Lense-Thirring effect (Muslimov \& Tsygan 1992; see also Muslimov \& Harding 1997).  Over much of the polar cap, particularly near the periphery, this produces ions and protons with only the moderately large Lorentz factors assumed here. 
Extensive past studies of pair creation based on the Lense-Thirring effect acceleration potential (see, for example, Hibschman \& Arons 2001, Harding \& Muslimov 2002) have shown that the formation of secondary electron-positron pairs by curvature radiation in a dipole field is possible only in a minority of pulsars.  (The necessary condition, drawn from Fig. 1 of Harding \& Muslimov, can be expressed as $B_{12}P^{-1.6} > 6.5$, where $B_{12}$ is the spin-down inferred polar-cap magnetic flux density in units of $10^{12}$ G. But we should note that many authors have attempted to obviate this by various ad hoc assumptions about deviations from a dipole field.)
Some pair formation by inverse Compton-scattered photons is always possible for the case of ${\bf \Omega}\cdot{\bf B} > 0$ neutron stars considered by these authors because the primary accelerated particles are electrons. But both papers comment on the small fluxes of secondary pairs that are predicted in many observable pulsars.

The essence of inverse Compton scattering is that a blackbody photon can take a substantial fraction of the energy of an outward accelerated electron. The initial perpendicular momentum  of the photon is small ($\ll 2m_{e}c$) so that conversion occurs at higher altitudes, above the polar-cap acceleration region, as a consequence of flux-line curvature.
The scattered photons, and hence the pairs, also have high energies.  But in the ${\bf \Omega}\cdot{\bf B} < 0$ case, there are no primary outward accelerated positrons.  Compton scattering by the inward accelerated photo-electrons merely produces a high-energy photon that enters the neutron-star surface.

High-energy electron or photon-induced showers do, in principle, produce a flux of backward-moving photons that enter the magnetosphere.
Possible sources are either purely electrodynamic, for example, multiple Compton scattering, or neutron capture, the neutron source being decay of the giant dipole state produced in shower development (see Jones 2010).  But positrons created close to the neutron-star surface are ineffective contributors to equation (4) because they are accelerated to energies close to the proton energy $(\gamma_{2} - 1)m_{p}$.
Pair production by the interaction of an energetic outward moving photon with the blackbody field is also a possible source.  The threshold condition is,
\begin{eqnarray}
k_{1}k_{2}(1 - \cos\theta) > 2(m_{e}c)^{2},
\end{eqnarray}
where $k_{1,2}$ are their momenta and $\theta$ the observer-frame angle between them, from which it follows that, for conservative blackbody temperatures and values of $\theta$, the energy of the photon and therefore of the pair must be high.

For these reasons, the flux of pairs that are produced in a normal ${\bf \Omega}\cdot{\bf B} < 0$ pulsar is expected to be small, but difficult to estimate reliably.  As a guide, the condition $C_{e} > C$ is an indication that the electron-positron component may affect the ion-proton Langmuir mode.  The ratio, from equations (5), is,
\begin{eqnarray}
\frac{C_{e}}{C} = \frac{\alpha_{e}}{\alpha_{p}}\frac{m_{p}}{m_{e}}
\frac{\gamma_{p}^{3}(\gamma_{m} + \gamma_{0})}{2\gamma_{0}^{2}\gamma_{m}^{2}},
\end{eqnarray}
from which it follows that even a small flux $\alpha_{e} \ll \alpha_{p}$ can have a significant effect on the dispersion relation provided its Lorentz factor limits, $\gamma_{0}$ and $\gamma_{m}$ remain not too large over path lengths of $\sim 10^{7}$ cm, long enough to give sufficient mode-amplitude gain.  This places a fairly stringent upper limit on the $E_{\parallel}$ component.  Investigation requires some trial solutions of equation (4) with $k_{\perp} = 0$.

\begin{figure}
\includegraphics[trim=20mm 25mm 10mm 25mm,clip,width=84mm]{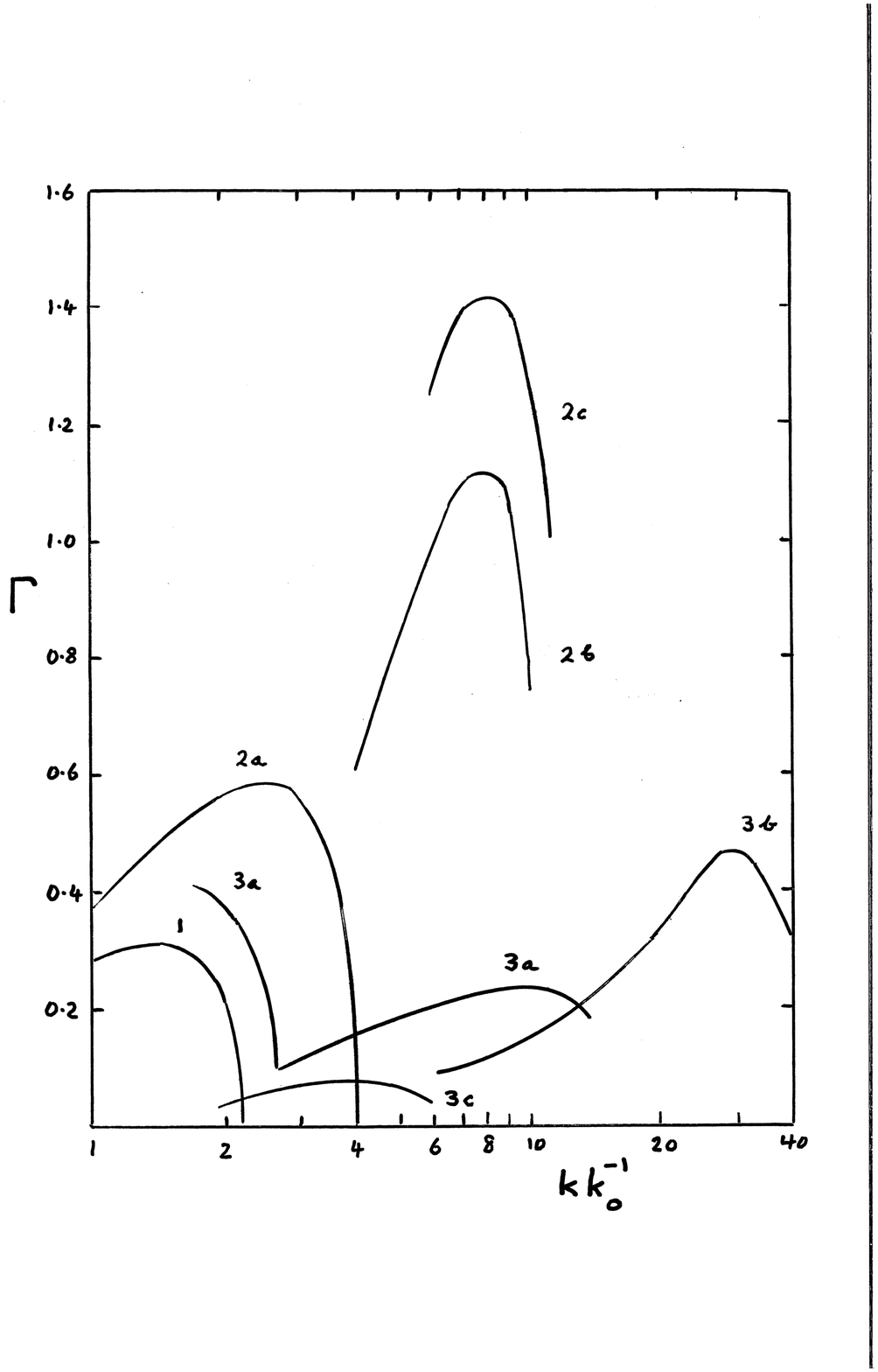}

\caption{Values of growth rates $\Gamma$ are shown in units of the fixed quantity $\omega_{1}^{*}$ as functions of the wavenumber $k$, which is in units of the constant $k_{0} = 2\gamma_{1}^{2}\omega_{1}^{*}$, evaluated for the fixed values $\gamma_{1} = 10$ and $\alpha_{1} =0.4$.  The curve labelled $1$ in the lower left-hand corner is the ion-proton growth rate in the absence of electrons and positrons.  The remaining curves are for the multi-component modes described as cases (ii) and (iii) in the text of Section 3. Curves $2a$ and $2b$ are for the partial overlap case with $\alpha_{e} = 0.05$ and $1.0$, respectively.  Curve $2c$ is for $\alpha_{e} = 1.0$ and shows the effect of departing from the fixed value of the proton current density to $\alpha_{2} = 0.1$ with $\alpha_{1} = 0.9$.  The consequent change in $\Gamma$ is not large. 
Curves $3a$ and $3b$ are for the no overlap case with $\alpha_{e} = 0.05$ and $1.0$ respectively, whilst $3c$ is as for $3a$ except that the electron-positron distribution limits $\gamma_{0,m}$ have each been increased by a factor of four.  The curves, which are incomplete, show only the interval of $k/k_{0}$ for which solutions have been obtained.  We refer to Section 3 for further details of parameter values.}
\end{figure}

We adopt fixed values, $\gamma_{1} = 10$, $\gamma_{2} =30$, and $\alpha_{1} = 0.4$, $\alpha_{2} = 0.6$ for the ion and proton Lorentz factors and current densities.
These are typical of previous model calculations (see Jones 2012b, 2013a).  The ions have some remaining electrons after the completion of acceleration and the Lorentz factors correspond to a mass to charge ratio
$A/Z_{\infty} = 3.22$.  With $C_{e} = 0$, the mode has a growth rate
$\Gamma = \omega_{1}^{*}{\rm Im}(s)$ which is finite for  wavenumbers $k/k_{0} \leq 2.20$ and is a linear function of wavenumber in the limit $k/k_{0} \rightarrow 0$.  A section of this
is shown in Fig. 1 (curve $1$), with $\Gamma$ in units of $\omega_{1}^{*}$, and has a maximum at $k/k_{0} \approx 1.4$.  We find that the growth rate is only a slowly varying function of the mass to charge and current density ratios.

The mode amplitude is $\propto\exp\Lambda(\eta)$ as a function of height above the polar cap, where the radius $\eta$ is in units of the neutron-star radius $R$.
An expression for the exponent $\Lambda$ has been given previously, under the assumption that the ion and proton Lorentz factors become only slowly-varying functions of altitude, and can be approximated by constants, well before the
emission region is reached (Jones 2012a,b).  It is,
\begin{eqnarray}
\Lambda = \left(\frac{-16\pi\alpha_{1}Z_{\infty}eR^{2}B\cos\psi}
{PAm_{p}c^{3}\gamma_{1}^{3}}\right)^{1/2}\left(1 - \eta^{-1/2}\right)
{\rm Im}(s),
\end{eqnarray}
assuming a dipole field, in which $\psi$ is the angle subtended by the magnetic and rotation axes.  This gives,
\begin{eqnarray}
\Lambda = 3.1\times 10^{2}\left(\frac{-B_{12}\cos\psi}{P}\right)^{1/2}\left(1 - \eta^{-1/2}\right){\rm Im}(s),
\end{eqnarray}
for the fixed parameters adopted here and $R = 1.2\times 10^{6}$ cm.  The dependence on $\gamma_{1}$ is interesting and shows that only moderately relativistic ions are necessary for high growth rates. The existence of lower values of $\gamma_{1}$ during some of the growth interval merely serves to increase $\Lambda$.
Thus ions or protons accelerated through a full polar-cap potential difference, unscreened by photo-electrons, will not usually produce useful growth rates. This expression remains valid for cases in which $C_{e} \neq 0$ on substitution of the appropriate value of
${\rm Im}(s)$. An exponent $\Lambda = 30$ was assumed to correspond with growth to the region of non-linearity and turbulence and previous papers adopted a fixed value ${\rm Im}(s) = 0.2$ for numerical estimates.

With a finite electron-positron flux, solutions of equation (4) depend on the sequence of the four singularities and it is convenient to distinguish three possible cases.

(i) 	The electron-positron distribution completely overlaps $s = -1$ and $-\mu$, that is, $\nu_{0} > 1$ and $\nu_{m} < \mu$, for which there can be either $4$ or $6$ real roots.  Graphical inspection shows that a guide to the existence of  $6$ real roots and hence no unstable mode is that $D_{zz}(s_{0}) > 0$, where
\begin{eqnarray}
s_{0} = - \left(\frac{\mu +C^{1/3}}{1 + C^{1/3}}\right)
\end{eqnarray}
is the extremum of $D_{zz}$ for the case that $C_{e} = 0$.  This has been confirmed by direct solution for $\gamma_{0} = 5$ and $\gamma_{m} = 100$, and we find that the ion-proton mode is extinguished by electron fluxes as small as $\alpha_{e} = 0.002$.

(ii)	Partial overlap with $\nu_{0} > 1$ and $\nu_{m} > \mu$ is, in general, a more likely case.  As with (i), there can be either $4$ or $6$ real roots.
The example of $\gamma_{0} = 20$, $\gamma_{m} = 500$ has been solved for various values of $\alpha_{e}$.  The intervals of $k/k_{0}$ are not complete, but establish the growth-rate maxima.  These are shown in Fig. 1.  For $\alpha_{e} = 0.05$ (curve $2a$), there is an unstable mode in the interval
$1 \leq k/k_{0} \leq 3.5$ with a maximum $\Gamma = 0.59$ at
$k/k_{0} \approx 2.5$.  There are $6$ real roots for $k/k_{0} \geq 4$.  With $\alpha_{e} = 1.0$ ($2b$), the mode is shown in the interval $4 \leq k/k_{0} \leq 10$ with a maximum $\Gamma = 1.12$ at $k/k_{0} \approx 8$.  The solution is one of $6$ real roots for $k/k_{0} \geq 12$.  For this latter set, we have investigated the effect of varying the initial fixed values so as to reduce the fraction of protons.  Curve $2c$ is for $\alpha_{e} = 1.0$ but with $\alpha_{2} = 0.1$ and $\alpha_{1} = 0.9$. The effect on $\Gamma$ is not large but indicates that two components is the optimum number for Langmuir-mode growth rate.  We have to emphasize here that this mode is not the canonical electron-positron mode which has previously been considered to be the coherent emission source because the momentum distribution $f(p)$ given in equation (3) certainly does not satisfy the Penrose condition (see Buschauer \& Benford 1977).  It is a mode involving all the charged particle types present.

(iii)	The final case of no overlap, $\nu_{0}, \nu_{m} < \mu$ is more complex, with the possibility of $2$, $4$ or $6$ real roots.  The growth rate is shown in Fig. 1 for $\gamma_{0} = 50$ and $\gamma_{m} = 500$ with $\alpha_{e} = 0.05$.  At small $k/k_{0}$ (curve $3a$), there are two unstable modes reducing to one with increasing $k/k_{0}$, having a maximum growth rate $\Gamma = 0.24$   at $k/k_{0} \approx 10$. There are 6 real roots at $k/k_{0} > 16$.  Also shown ($3c$) is the effect of increasing electron-positron energies to $\gamma_{0} = 200$ and $\gamma_{m} = 2000$, which reduces the growth rate significantly.    The case $\gamma_{0} = 50$, $\gamma_{m} =500$, and $\alpha_{e} = 1.0$ is also shown on the same diagram ($3b$).  Here, the unstable mode persists to $k/k_{0} > 40$ and has a maximum growth rate $\Gamma = 0.47$ at $k/k_{0} \approx 30$.  Its angular frequency is
\begin{eqnarray}
\omega = k - \frac{k_{0}}{2\gamma_{1}^{2}}\left(\frac{k}{k_{0}} -1
 -{\rm Re}(s)\right),
\end{eqnarray}
and, with ${\rm Re}(s) = 26.12$ at the maximum, it is subluminal.  This is very similar to case (ii).

Quantitatively, the results shown in Fig. 1 should not be given too much weight.  It was remarked earlier that there appear to be no published calculations of the electron-positron energy distribution in the classical region of single-photon magnetic conversion, where Landau quantum numbers can be very large.  Numerical results obtained here are likely to be sensitive to the specific form of the distribution and the one adopted here may not be realistic.

The Langmuir modes exist, in principle, for a finite interval of wavenumber.  But we envisage that following the initial fluctuation, the mode will rapidly develop towards the nearest growth-rate maximum.  The real polar-cap plasma is far from the infinite one-dimensional ideal owing to finite transverse dimensions and parameter gradients.  The longitudinal mode itself does not couple with radiative modes. Although Asseo et al (1990) recognized that the general quasi-longitudinal mode ($k_{\perp} \neq 0$) does  couple it has not been considered here in detail, in part because there seems to be no reason why $k_{\perp}$ should be a constant of motion.  In fact all the parameters of the mode change (in essence, adiabatically) as its amplitude increases with altitude above the polar cap. The effect of flux-line curvature on growth rates is negligible, as can be verified by reference to the paper of Asseo et al. Following our comments in the first paragraph of Section 1, we see no reason why a turbulent distribution of charge and current fluctuations, whether formed by longitudinal or quasi-longitudinal modes and given the parameter spatial gradients that exist and the lack of spatial symmetry, should not couple directly with radiative modes. 

\section{Conclusions}

Earlier papers (see Jones 2014b) developed the case for the ion-proton mode as the source of plasma turbulence decaying through coherent emission at radio frequencies.  But owing to the small electron-proton mass ratio, even a small electron-positron flux can modify the character of the dispersion relation, and it is not possible to exclude such a flux, which may arise from several different processes.  The present paper describes a study of the multi-component Langmuir modes that result. Of necessity, we have been obliged to make a specific assumption about the form of the electron-positron momentum distribution present in equation (2).  Although this may be to some extent in error, we suggest that the general character of the multi-component mode would probably survive substitution of the correct distribution.

The multi-component mode complicates the physics of the pulsar magnetosphere because it introduces uncertain factors additional to those, principally surface nuclei atomic number and neutron-star surface temperature, that affect the ion-proton mode considered previously.  But this appears an advantage rather than the reverse because the single most obvious pulsar characteristic, apart from the coherent emission, is the number of them that can be described only as sui generis.  Broad distributions of parameters are the norm even for selected sets, such as spectral indices (see Malofeev et al 1994) or the radio-frequency power emitted per primary beam particle (Jones 2014b).  With this in mind, the principal results of Section 3 can be summarized as follows.

For a momentum distribution such that the $\delta$-function ion and proton Lorentz factors are completely overlapped by those of the electrons and positrons, the plasma is stabilized and there is no Langmuir mode with finite growth rate.  Under the specific assumption made here, in equation (3), this is true even for electron-positron fluxes two orders of magnitude smaller than the Goldreich-Julian flux.  Usually, this condition will be satisfied for polar-cap magnetic flux densities of the order of $10^{14}$ G, as in magnetars, where copious production of low-energy pairs is expected through the processes described by Thompson (2008). In these objects, radio-frequency emission with spectral properties similar to those of normal and millisecond pulsars should not, in general, be seen.
This does not appear to be inconsistent with present observations.  Although Malofeev et al (2005) have published flux densities at 67, 87 and 111 MHz, with large errors, for the anomalous X-ray pulsar J1308+2127, the ATNF pulsar catalogue (Manchester et al 2005) has no entries at the standard higher frequencies for this object. It also makes no reference to the other objects mentioned by these authors. It appears that the emission from J1810-197 is well-established (Camilo et al 2007), but its spectrum extends to 144 GHz and the spectral index is quite different from that of the normal and millisecond pulsars.It also has other properties not observed in the general pulsar population (see Serylak et al 2009). 

In normal and millisecond pulsars, the multi-component mode has wavenumber  and growth rate dependent on the electron-positron flux and we have argued that, for the ${\bf \Omega}\cdot{\bf B} < 0$ case, this will be one or more orders of magnitude smaller than the Goldreich-Julian flux.  There must be some small reverse flux of electrons accelerated inward which, with photo-electrons, reduces the ${\bf E_{\parallel}}$ field component.  These have been explicitly neglected in equations (3) and (4), but their contribution to the dielectric tensor is not large and the character of the solutions described in Section 3 should not be affected.  Mode angular frequencies are variable but growth rates can be much larger than that for the ion-proton mode.  This variability is by no means inconsistent with the observed properties of the pulsar population, as we have emphasized above.

A factor not so far mentioned is the capacity of the different components in the multi-component mode to store energy.  This is of no consequence at the extremely small amplitudes of the initial stages of growth. But later growth immediately prior to non-linearity and the formation of plasma turbulence might  be influenced by the very large changes in electron-positron Lorentz factors that would be expected.  This, with the nature of the plasma turbulence and its decay, are problems that require further work.

A remaining question concerns ${\bf \Omega}\cdot{\bf B} > 0$ neutron stars.
In these, the polar-cap plasma has only two components, electrons and positrons.  It appears difficult to generate plasma turbulence through Langmuir modes in this case and the Penrose condition may not even be satisfied.  There must be doubts that these objects have any observable polar-cap coherent emission except possibly when young.  Otherwise, incoherent emission from the vicinity of the light cylinder may be all that is observable.

\bsp

\label{lastpage}

\end{document}